\documentclass[usenatbib]{mn2e}
\usepackage{multirow}
\usepackage{rotating}
\usepackage{colortbl}
\usepackage{color}
\usepackage{xcolor}
\usepackage[fleqn]{amsmath}
\usepackage{amssymb}
\usepackage{amsfonts}
\usepackage{verbatim}
\usepackage{scalefnt}
\usepackage{lscape}
\usepackage{booktabs}
\usepackage{multicol}

\usepackage[bookmarks=False,bookmarksnumbered,colorlinks=true, citecolor=blue, linkcolor=cyan]{hyperref}

\newcommand{\beq}{\begin{equation}}
\newcommand{\eeq}{\end{equation}}

\newcommand{\msun}{\,{\rm M_\odot}}

\title[MBHs in $z>9$ JWST galaxies ]
{What if young $z>9$ JWST galaxies hosted massive black holes?}

\author[Volonteri et al.]{Marta Volonteri,$^{1}$\thanks{E-mail: martav@iap.fr (MV)}
M\'elanie Habouzit,$^{2,3}$
Monica Colpi$^{4,5}$
\\
$^{1}$Institut d'Astrophysique de Paris, Sorbonne Universit\'e, CNRS, UMR 7095, 98 bis bd Arago, 75014 Paris, France\\
$^{2}$Zentrum f\"ur Astronomie der Universit\"at Heidelberg, ITA, Albert-Ueberle-Str. 2, D-69120 Heidelberg, Germany\\
$^{3}$Max-Planck-Institut f\"ur Astronomie, K\"onigstuhl 17, D-69117 Heidelberg, Germany\\
$^{4}$Dipartimento di Fisica G. Occhialini, Universita di Milano-Bicocca, Piazza della Scienza 3, I-20126 Milano, Italy\\
$^{5}$INFN, Sezione di Milano-Bicocca, Piazza della Scienza 3, I-20126 Milano, Italy\\
}

\date{2022}

\begin{document}
\maketitle

\begin{abstract}
JWST is discovering star forming `candidate' galaxies with photometric redshifts $z>9$ and little attenuation. We model presumptive massive black holes (MBHs) in such galaxies and find that their unobscured emission is fainter than the galaxy starlight in JWST filters, and difficult to be detected via color-color selection, and X-ray and radio observations. Only MBHs  overmassive relative to expected galaxy scaling relations, accreting at high Eddington rates, would be detectable. Their discovery would point to the presence of heavy MBH seeds, but care is needed to exclude the existence of lighter seeds  as only overmassive MBHs are detectable in this type of galaxies.  Conversely, if no overmassive MBHs are hosted in these galaxies, either there are no heavy seeds or they are rare. The most massive/highest redshift candidate galaxies can attain stellar masses in excess of $5\times 10^{10}\msun$ by $z\sim 6$ if they grow along the SFR-mass sequence, and can nurse a MBH growing from $\sim 10^5 \msun$ up to $>3\times 10^7\msun$ by $z\sim 6$, to become hosts of some $z > 6$ quasars. Candidate galaxies of  $\log( M_{\rm gal}/\msun)\sim 8$ can not grow their putative seeds fast, unless seeds are $\gtrsim 10^6\msun$.  The number density of the JWST candidate galaxies far outnumbers that of the highest-$z$ quasar hosts and this allows for about only 1 bright $z\sim 6-7$ quasar every 1000 of these galaxies.

\end{abstract}

\section{Introduction}\label{Intro}

A new frontier on high-redshift galaxy studies has been opened with the launch of JWST. Within a few months of operation a wealth of new galaxy candidates at $z>9$ has been identified from photometric redshifts \citep{2022arXiv220709434N,2022arXiv220709436C,2022arXiv220712338A,2022arXiv220711558Y,2022arXiv220711217A,2022arXiv220711135L,Harikane2022,2022arXiv220804292D,2022arXiv221001777B,2022arXiv220712446L,2023MNRAS.519..157W}, and some with spectroscopic confirmation \citep{2022arXiv220710034S,2022arXiv221015639R}. 

Most of the $z>9$ candidates are young, star forming and appear to have little or no dust \citep{2022arXiv220800720F}, and they are suggested to have been picked up in observations exactly for these reasons \citep{2022arXiv220714808M}. There has been much discussion on whether these galaxies are expected in theoretical models, and whether they challenge the galaxy formation paradigm \citep[see, e.g.,][for a discussion]{2022arXiv220712474F}. In general their numbers are higher than expected \citep{2022arXiv221105792F}, but for the majority of cases the build-up of the stellar masses is not incompatible with models \citep{2022arXiv220801611B,2022arXiv221010066K}, and while some `all-purpose' simulations struggle to reproduce the observations at $z>12$, simulations dedicated to the high-redshift Universe fare better \citep{2017ApJ...836...16D,2022MNRAS.tmp.3128W}. Inclusion of different generations of stars and detailed dust treatment \citep[e.g.,][]{2020MNRAS.494.1071G,2022ApJ...936...45H} are likely key in improving the understanding of these galaxies.\\

In general, we consider here that a MBH could be lurking in a galaxy without dominating the emission at restframe optical/UV wavelengths. When the accretion luminosity is higher than the luminosity due to star formation, an AGN can be identified at these wavelengths by color selection \citep[e.g.,][]{2001AJ....121...31F}, or by emission line diagnostics when spectroscopy is available \citep{1981PASP...93....5B,2022arXiv221113648V}. X-ray and radio can also be used to distinguish star formation- and accretion-powered sources when one significantly dominates over the other. Finally, if a source is very compact, compatible with the point spread function of a high-angular resolution instrument, one could argue that the lack of extended emission is a signature of an accretion-dominated source. In the case of faint and small sources, such as high-redshift galaxies or also low-mass MBHs in low-redshift dwarfs, it is generally difficult to uniquely determine if an AGN is present. Often, multi-wavelength analysis is required for confirmation, with many sources remaining ``candidates'', as discussed in \cite{2020ARA&A..58..257G}.

Most of these candidate galaxies are presented as being dominated by star formation, without the presence of an Active Galactic Nucleus (AGN), with the exception of GL-z12, which is a candidate AGN \citep{2022arXiv220813582O} based on its compact size. The templates used for determining the physical properties, such as stellar mass, star formation rate (SFR) and age generally assume the absence of an AGN.  A separate question, which we do not address here, is whether an AGN template could be an alternative to a stellar template to explain the photometric properties of these sources. \cite{2022MNRAS.514L...6P} show an example of $z\sim 13$ galaxies that could be powered either by star formation or by a quasar. \\

In this paper we consider the physical properties presented in the discovery papers and ask what type of massive black holes (MBHs) and AGN could be hidden there, and what type of MBHs and AGN could be detected in galaxies with the redshift, stellar mass, star formation rates typical of these galaxies. We then explore the implications for MBH seed models and for understanding the build-up of  $z>6$ quasars. 

\section{What massive black holes could be hidden in these galaxies?}\label{multilambda}

\begin{table*}
\begin{tabular}{llllllll}
ID           & $\log(M_{\rm gal}/\msun)$ & SFR ($\msun$/yr) & $z_{\rm phot}$ & MUV   & F356Wapp & F444Wapp & Comments            \\
\hline
GL-z10       & 9.60  & 10.00          & 10.40  & -21.00 & 26.50     & 0.00     & N22$^*$ (GL-z9-1 in H22)               \\
GL-z12       & 9.10  & 6.00           & 12.30  & -20.70 & 0.00      & 0.00     & N22$^*$               \\
SMACS\_z10a  & 8.86  & 0.01           & 9.77   & -18.77 & 0.00      & 0.00     & Fu22                     \\
SMACS\_z10b  & 10.21 & 0.04           & 9.03   & -20.78 & 0.00      & 0.00     & Fu22                     \\
SMACS\_z10c  & 9.72  & 0.47           & 9.78   & -20.19 & 0.00      & 0.00     & Fu22                     \\
SMACS\_z10d  & 6.95  & 3.47           & 9.32   & -19.76 & 0.00      & 0.00     & Fu22                     \\
SMACS\_z10e  & 6.87  & 14.45          & 10.88  & -18.91 & 0.00      & 0.00     & Fu22                     \\
SMACS\_z11a  & 6.46  & 5.89           & 11.08  & -18.55 & 0.00      & 0.00     & Fu22                     \\
SMACS\_z12a  & 8.27  & 0.05           & 12.16  & -19.75 & 27.70     & 0.00     & Fu22                     \\
SMACS\_z12b  & 8.26  & 0.10           & 12.27  & -20.01 & 28.20     & 0.00     & Fu22                     \\
SMACS\_z16a  & 8.02  & 16.60          & 15.93  & -20.59 & 27.80     & 0.00     & Fu22                     \\
SMACS\_z16b  & 7.89  & 57.54          & 15.25  & -20.96 & 0.00      & 0.00     & Fu22                     \\
Maisie's       & 8.45  & 4.10           & 14.30  & -20.30 & 28.05     & 28.28    & Fi22 (CR2-z12-1 in H22)                  \\
WHL0137-3407 & 8.78  & 7.30           & 10.70  & 0.00   & 27.13     & 27.15    & B22                   \\
WHL0137-5021 & 8.53  & 5.10           & 12.80  & 0.00   & 28.12     & 27.94    & B22                   \\
WHL0137-5124 & 8.65  & 6.90           & 12.70  & 0.00   & 28.02     & 27.99    & B22                   \\
WHL0137-5330 & 8.77  & 6.40           & 10.00  & 0.00   & 27.45     & 27.27    & B22                   \\
WHL0137-5347 & 9.01  & 14.60          & 10.20  & 0.00   & 26.60     & 26.51    & B22                   \\
WHL0137-8737 & 8.46  & 6.00           & 9.20   & 0.00   & 27.20     & 27.40    & B22                   \\
JD1          & 7.90  & 0.13           & 9.76   & -17.45 & 27.81     & 27.82    & RB22                  \\
GL-z9-1      & 9.15  & 27.00          & 10.68  & -20.20 & 26.50     & 0.00     & H22 (GL-z10 in N22)  \\
CR2-z12-1    & 8.38  & 3.40           & 11.88  & -19.70 & 27.90     & 0.00     & H22 (Maisie's in Fi22)   \\
GL-z12-1     & 8.56  & 3.00           & 12.22  & -20.80 & 27.00     & 0.00     &  H22                    \\
S5-z12-1     & 8.08  & 2.20           & 13.72  & -20.30 & 27.60     & 0.00     &  H22                     \\
CR2-z17-1    & 8.77  & 9.10           & 16.45  & -21.90 & 26.30     & 0.00     &  H22                     \\
S5-z17-1     & 8.84  & 9.70           & 16.66  & -21.60 & 26.60     & 0.00     &  H22                    \\
\hline
\end{tabular}
\caption{List of candidate galaxies considered in this paper. N22=\citet{2022arXiv220709434N}; Fu22=\citet{2022arXiv220805473F}; Fi22=\citet{2022arXiv220712474F}; B22=\citet{2022arXiv221001777B}; RB22=\citet{2022arXiv221015639R}. When the photometry is not listed in the discovery papers, we obtained it from \citet{Harikane2022}.  The typical statistical 1-$\sigma$ uncertainties in stellar masses are generally between 0.1 and 1~dex, but systematic uncertainties can be larger. $^*$Average of the two $z_{\rm phot}$ in the paper. Two galaxies appear in different papers with somewhat different inferred physical properties. }
\label{TableGal}
\end{table*}

\begin{figure*}
	\includegraphics[width=\textwidth]{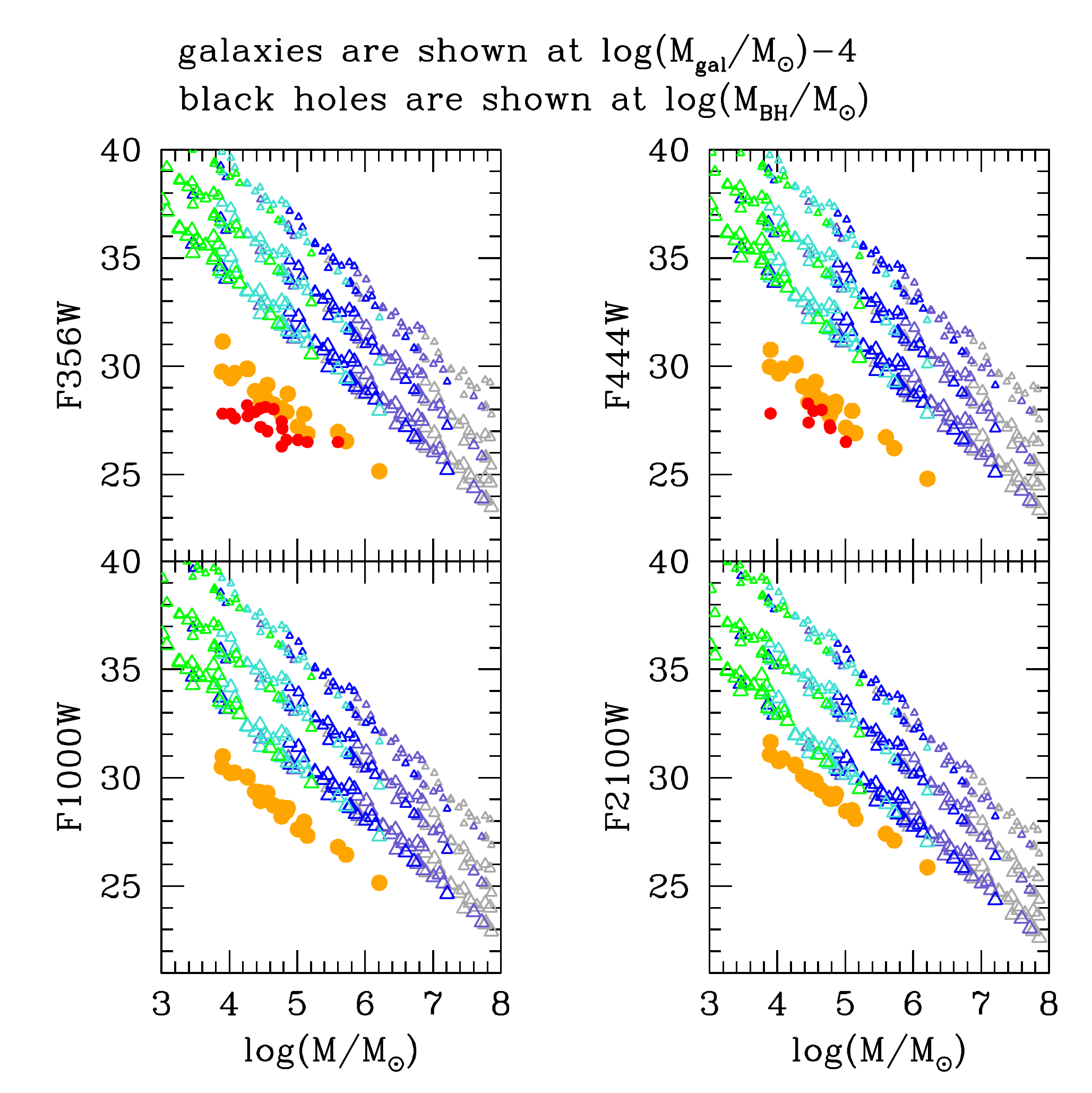}
    \caption{Comparison of AGN and galaxy luminosity using our models: apparent magnitude of AGN and galaxies.  
    Triangles show the AGN magnitudes corresponding to MBHs following different scaling relations with the host galaxies, and accreting at fractions of the Eddington limit. Green: MBHs with mass $\log(M_{\rm gal}/\msun)-5$; turquoise: MBHs with mass $\log(M_{\rm gal}/\msun)-4$; blue: MBHs with mass $\log(M_{\rm gal}/\msun)-3$; slate grey: MBHs with mass $\log(M_{\rm gal}/\msun)-2$; grey: MBHs with mass $\log(M_{\rm gal}/\msun)-1$. The size of the symbol scales with the Eddington ratio: small for $\log(f_{\rm Edd})=-2$, medium for $\log(f_{\rm Edd})=-1$, large for $\log(f_{\rm Edd})=0$.    
    Red dots: F356W or F444W  from observational references (only a fraction of the candidates have published photometry, therefore some of the sources are not shown). Photometric errors are not shown for clarity but they are less than 10 per cent. Orange dots: galaxy apparent magnitude from our models. Galaxies are shown at the mass corresponding to $\log(M_{\rm gal}/\msun)-4$. The simple galaxy model we use produces reasonable results. When comparing AGN and galaxy magnitudes we see that AGN can be brighter than the host only when they are overmassive with respect to the nominal relation and they have high Eddington ratios.}
    \label{fig:compareL}
\end{figure*}

\begin{figure}
\includegraphics[width=\columnwidth]{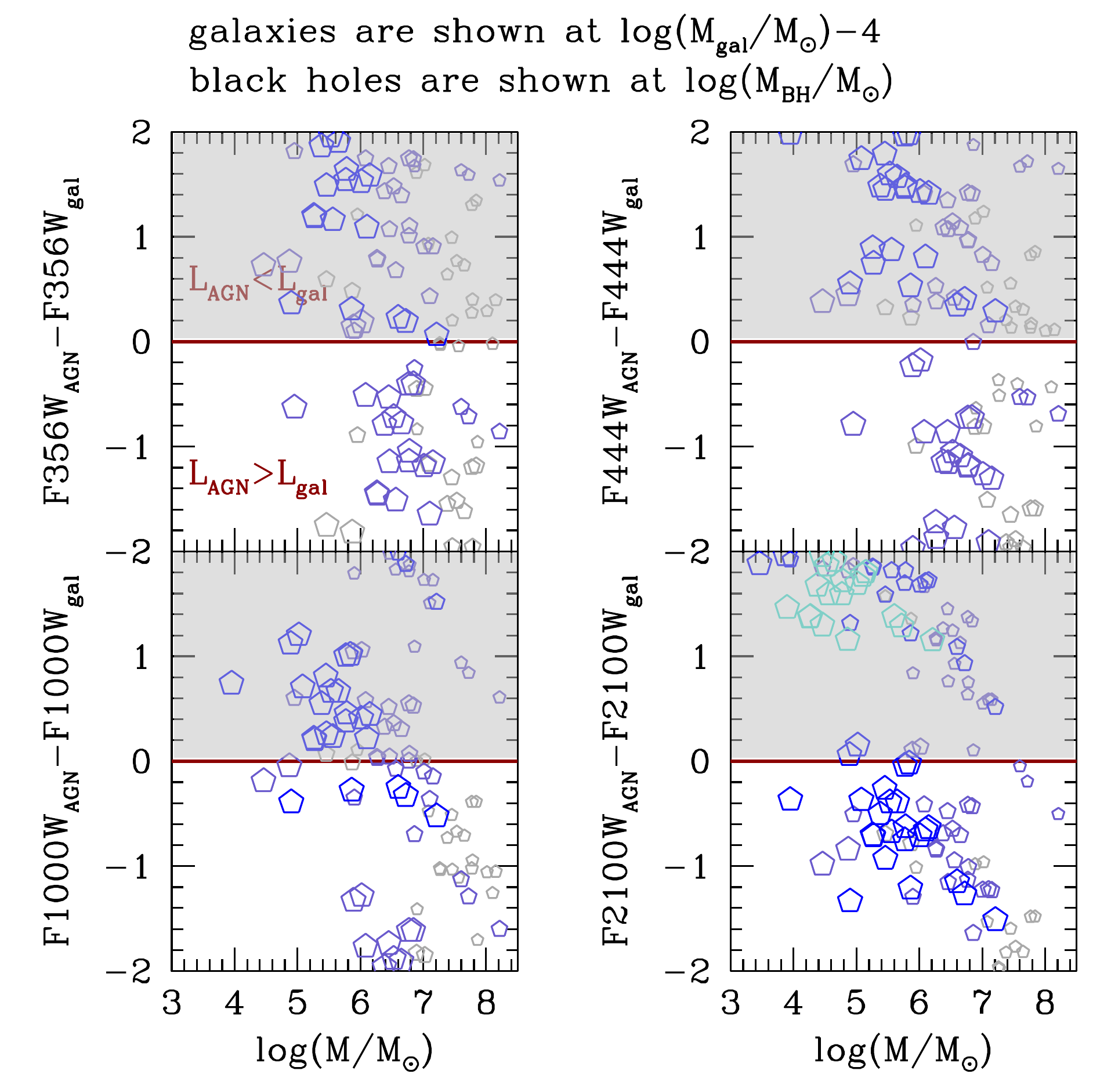}
\caption{Difference between the AGN and galaxy magnitude (shown only for magnitude differences less than $\pm2$). Colors as in Fig.~\ref{fig:compareL} and the symbol size also scales with Eddington ratio as described in that Figure. The grey shaded area shows the region where the AGN is fainter than the stellar component. 
At most JWST wavelengths only MBHs with mass larger than $\log(M_{\rm gal}/\msun)-2$ can outshine their host galaxy, whereas MBHs with mass $\log(M_{\rm gal}/\msun)-2$ and $\log(f_{\rm Edd})=0$ can outshine the galaxy at the longest JWST wavelengths. MBHs on the nominal relation are always fainter than their host galaxy. }
    \label{fig:compareLzoom}
\end{figure}

\begin{figure}
	\includegraphics[width=\columnwidth]{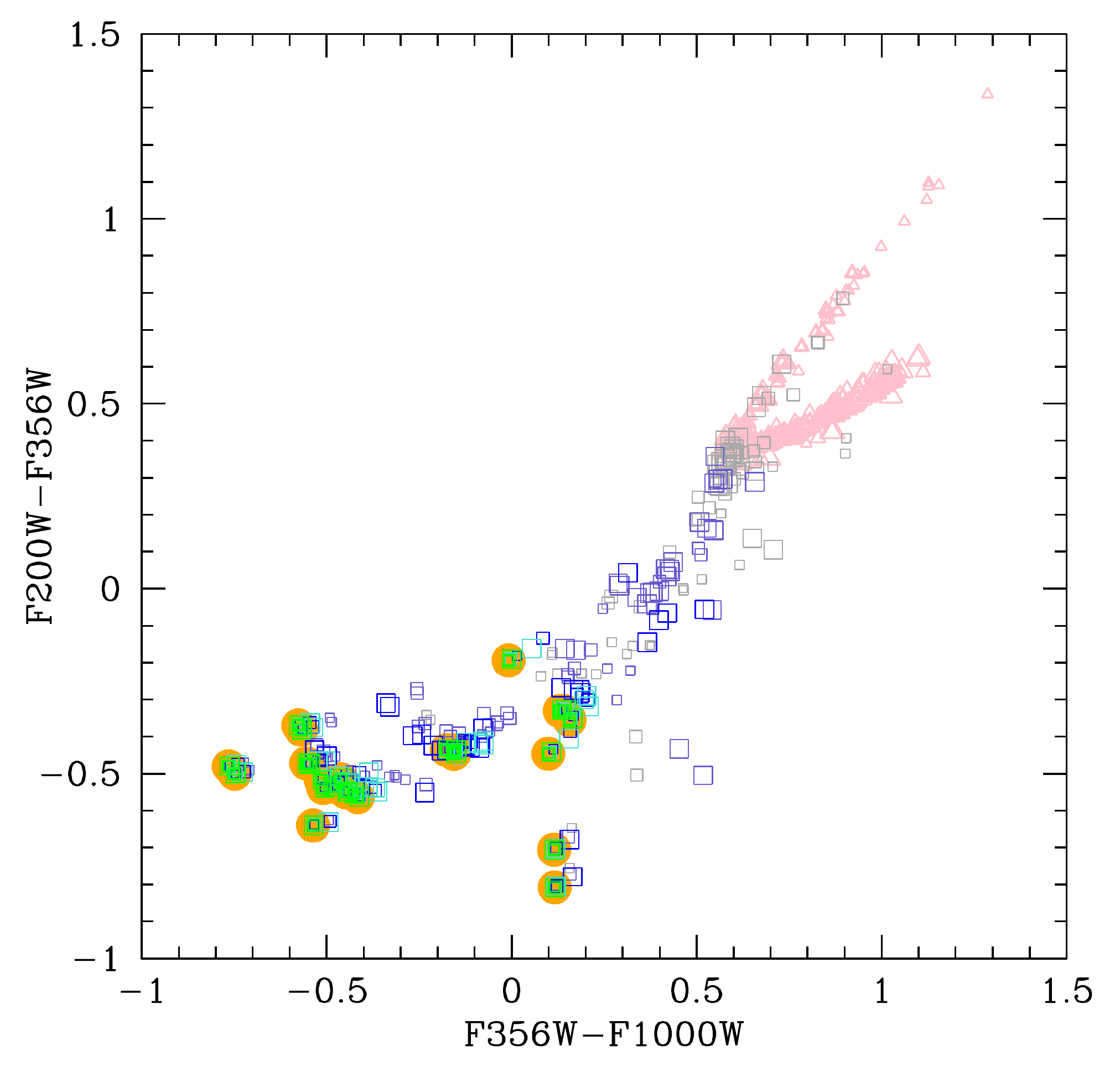}
    \caption{Pink triangles: AGN-only colors; to make the figure legible we do not differentiate the color of MBHs with different MBH-to-galaxy mass ratios.  The size of the symbol scales with the Eddington ratio: small for $\log(f_{\rm Edd})=-2$, medium for $\log(f_{\rm Edd})=-1$, large for $\log(f_{\rm Edd})=0$.  AGN are mostly filling the top-right corner, being `redder' than the star forming galaxies of the simulated sample.  Orange dots: galaxy-only colors. Galaxies instead tend to populate the lower-left corner and are `bluer' in the JWST filters.  Squares: AGN+galaxy colors.  Green: MBHs with mass $\log(M_{\rm gal}/\msun)-5$; turquoise: MBHs with mass $\log(M_{\rm gal}/\msun)-4$; blue: MBHs with mass $\log(M_{\rm gal}/\msun)-3$; slate grey: MBHs with mass $\log(M_{\rm gal}/\msun)-2$; grey: MBHs with mass $\log(M_{\rm gal}/\msun)-1$.   Only substantial and rapidly accreting MBHs can be distinguished.}
    \label{fig:combcolorstest}
\end{figure}

\subsection{Galaxy and AGN multi-wavelength modelling}

The properties of the candidate galaxies are reported in Table~\ref{TableGal}. We have used for our models the best-fit values presented in the papers; the statistical 1-$\sigma$ uncertainties in stellar masses are generally between 0.1 and 1~dex, although for some galaxies it can be larger (e.g., CR2-z12-1, S5-z17-1). The uncertainties in the SFRs have a large spread, for instance they are generally less than 50 per cent for the galaxies presented in \citet{2022arXiv221001777B}, while they can reach 100 per cent in \cite{2022arXiv220805473F} and \cite{Harikane2022}. In terms of systematic uncertainties, in Table~\ref{TableGal} we include two galaxies that have been reported by different papers, GL-z10/GL-z9-1 and CR2-z12-1/Maisie's, to give an idea of the interpublication scatter, which can reach high values. For instance SMACS\_z10e and SMACS\_z11a differ in mass by about two orders of magnitude between \cite{2022arXiv220805473F} and \cite{2022arXiv220712338A}, with the values from the former study being preferred (H. Atek, private communication). These systematic differences are generally ascribed to choices on how to model the star formation history \citep{2023MNRAS.519..157W}.

 Based on the mass, star formation rate (SFR), and photometric redshift of the observed galaxy candidates, we run a grid of galaxy and MBH models to obtain their multi-wavelength properties in different JWST filters, and in the X-ray and radio bands. The galaxy and AGN spectral energy distributions (SEDs) are described and shown in \cite{2017ApJ...849..155V}: from these SEDs we calculate galaxy and AGN absolute and apparent magnitudes\footnote{For both galaxies and AGN we integrate the SED (convolved with the filter response) only redwards of 912 \AA. This differs with respect to \cite{2017ApJ...849..155V}, where magnitudes were calculated at the central wavelength of the filter, without convolution.}, as well as the expected  AGN X-ray fluxes, which we use in turn to estimate the radio fluxes. The galaxy X-ray emission is based on the X-ray binary population, and therefore depends on mass and SFR. 

 For each galaxy we model a grid of MBH masses and Eddington ratios, assuming a 10 per cent radiative efficiency. MBH masses, $M_{\rm BH},$  go from $\log(M_{\rm gal}/\msun)-5$ to $\log(M_{\rm gal}/\msun)-1$ in steps of 1 dex: this allows us to test the presence of undermassive and overmassive MBHs with respect to the nominal relation, that we consider to be $\log(M_{\rm gal}/\msun)-4$ at these redshifts and for these galaxy masses \citep{2021arXiv210510474Z}.  We also consider that `normal' MBHs can be within 1 dex of the nominal relation, thus truly overmassive MBHs are those with mass $>\log(M_{\rm gal}/\msun)-2$. Eddington ratios vary in the interval between $\log(f_{\rm Edd})=-2$ to $\log(f_{\rm Edd})=0$ in steps of 1 dex. We do not consider lower and higher accretion rates because our model is based on standard radiatively efficient thin discs \citep{1973A&A....24..337S}.

\subsubsection{Galaxies}
Galaxy spectra are based on Bruzual \& Charlot models \citep[][version 2016]{2003MNRAS.344.1000B}, adopting a Salpeter initial mass function. We assume constant star formation histories and map stellar mass to age through the SFR\footnote{While ages are estimated in many of the discovery papers, their definition is not consistent from one paper to another. This is the reason why we prefer to estimate the age based on the constant star formation histories we adopt for the stellar populations.}: $\rm{age}=\rm{min}(M_{\rm gal}/{\rm SFR},t_{\rm zphot})$, where $t_{\rm zphot}$ is the age of the Universe at $z_{\rm phot}$. We then assign to a galaxy a metallicity bin, either $10^{-2.3}\, Z_\odot$, $10^{-0.7}\, Z_\odot$ or solar, applying the $z=9$ mass-metallicity relation of  \cite{2022arXiv221016750N}\footnote{This is at variance with \cite{2017ApJ...849..155V}, where the mass-metallicity relation at $z=6$ from \cite{2016MNRAS.456.2140M} was used.}. In this paper we consider only unattenuated spectra, since for these candidates dust is expected to be minimal \citep{2022arXiv220800720F,2022arXiv220805473F} and indeed we find that the agreement in the photometry between model and observations worsens if we include dust using the model adopted in \cite{2017ApJ...849..155V}. We stress that the galaxy SEDs use very simple approaches, but they give reasonable results when compared to observations (see Appendix~\ref{App:phot}). 

Young star forming galaxies host populations of bright X-ray binaries. Empirical models for galaxy-wide X-ray emission from X-ray binaries are based either on the observations of nearby galaxies \citep[e.g. $<50 \rm \, Mpc$,][]{2019ApJS..243....3L} or on stacked galaxies observed at higher redshift \citep[$z<5$]{Fornasini2018}. We calculate the combined luminosity  in the  galaxies  using the relations of \cite{Fornasini2018}; the total emission from galaxies at higher redshift (e.g., the $z>9$ range considered in this paper) is still unknown, and could be even higher due to more abundant high-mass binaries. 

Radio emission from star-forming regions could be of the same order of magnitude as that of AGN powered by MBHs with mass $\lesssim 10^7 \msun$ based on \cite{2003ApJ...586..794B}. We chose not to include it explicitly in the analysis here because \cite{2003ApJ...586..794B} fit SFR as a function of radio luminosity, rather than vice versa, which is what we need. Furthermore, their study does not extend to high redhsift. An order of magnitude estimate obtained inverting the relation between SFR and radio luminosity in \cite{2003ApJ...586..794B} suggests that the SFR-driven radio luminosity remains below the sensitivity of planned surveys, although it could be higher than the radio emission from AGN for MBHs with mass $<10^7 \msun$.

\subsubsection{AGN}

AGN spectra (continuum only\footnote{We note that emission lines, which we don't include, could increase the flux in some bands \citep{2013ApJ...763..129S}.}) are described by the following equation:
\begin{equation}
f_\nu = {\cal N}\left(\nu^{\alpha_{\rm UV}} e^{-\frac{h\nu}{kT_{\rm BB}}}e^{-\frac{kT_{\rm IR}}{h\nu}}+a \nu^{\alpha_X}\right),
\label{AGN_SED}
\end{equation}
with $\alpha_{\rm UV}$=0.5 and $\alpha_X=1$, $k T_{\rm IR}=0.01$ Ryd. They are based in optical/near-IR on the Shakura-Sunyaev solution \citep{1973A&A....24..337S}, following \cite{2016ApJ...833..266T}. The model is calibrated in X-rays using results from the physical models developed by \cite{2012MNRAS.420.1848D}: the normalization $a$ is obtained through $\alpha_{\rm OX}$,the exponent of a power-law connecting the continuum between 2 keV and 2500 \AA, fitting the dependence on MBH mass  and Eddington ratio using the results in \cite{Dong2012}.  The last term in Equation~\ref{AGN_SED} is set to zero below 1.36 eV (912 nm). The global normalization ${\cal N}$ is obtained by requiring that the bolometric luminosity matches $L_{\rm bol}=1.26\times 10^{38} \rm{erg \, s^{-1}} f_{\rm Edd} M_{\rm BH}$. Also in this case we consider only unattenuated spectra. X-ray luminosity is calculated in the [2-10]~keV range (observer frame)

The radio luminosity is calculated via the fundamental plane of black hole accretion, an empirical correlation between the MBH mass, the 5 GHz radio and 2-10 keV X-ray power-law continuum luminosities \citep[FP,][note that this is the core radio luminosity and not the total luminosity including extended jets]{2019ApJ...871...80G}. We also include a variant (enhanced FP) where we increase the radio luminosity adding a boost in log space, with equal probability between 0 and 4.  This is motivated by low-mass MBHs being in some cases offset from the fundamental plane, i.e., having a radio luminosity up to 4 orders of  magnitude larger than predicted by the FP \citep{2022arXiv220909890G,2014ApJ...788L..22G}. Radio luminosity is calculated at 2~GHz (observer frame), assuming a power-law spectrum with index $-0.7$ \citep{2014ApJ...788L..22G}. 

\subsection{Properties or detections of MBHs in JWST bands}

In Fig.~\ref{fig:compareL} we show a comparison of the galaxy and AGN properties in JWST bands, computing galaxy and AGN apparent magnitudes from our model.  Since most of these galaxies are described in their discovery papers as not dominated by AGN \citep{Harikane2022,2022arXiv220714265T}, this analysis  gives limits  to  both the MBH mass and accretion rate. In Fig.~\ref{fig:compareL} the mass of galaxies  $M_{\rm gal}$ on the $x-$axis is shifted so that orange and red dots, representing the galaxies of our sample, are placed at $\log(M_{\rm gal}/\msun)-4$. MBHs are placed at the mass-scale in the grid we created (going from $\log(M_{\rm gal}/\msun)-4$ to $\log(M_{\rm gal}/\msun)-1$). 

In the two top panels we compare the observed magnitudes of the candidate galaxies of Table \ref{TableGal} (red dots in the filters F356W and F444W, ascribed to stellar populations according to the discovery papers) to our modelled ones (orange dots), if we only include the stellar contribution calculated using the tabulated $z_{\rm phot}$, SFR and stellar mass. 
Although we do not have a one-to-one match, observed and modelled magnitudes are very similar, reassuring us of our models being acceptable. 

Fig.~\ref{fig:compareLzoom} zooms into the magnitude difference between simulated MBH and galaxy SEDs, highlighting more clearly the relation between the starlight and AGN. We note that of course the presence of an unaccounted for AGN in the galaxies would modify the estimated masses and star formation rates (and perhaps even photometric redshifts), but if we take face value the stance that the measured flux from these galaxies is generated fully by stellar population, MBHs on the nominal relation between the galaxy and the MBH could be hidden there, and remain invisible, since their contribution to the flux is minimal at all explored JWST bands. An example is GL-z12, where \cite{2022arXiv220813582O} explore the possible presence of an AGN: in the first place, the galaxy mass and SFR have to be re-evaluated if the AGN produces a fraction of the light, and in the second place, they also concord with our suggestion that the MBH should have mass of the order of $10^6 \msun$ to be visible, for a galaxy with mass $2\times 10^8 \msun$, i.e., the MBH would have to be `overmassive'. HD1 and HD2 are two sources at $z\sim 13$ that have been proposed to be either star-forming galaxies \citep{2022ApJ...929....1H} or quasars \citep{2022MNRAS.514L...6P}. In  the latter possibility they would also be powered by a MBH on the overmassive side, with an inferred MBH mass of $\sim 10^8 \msun$ (assuming Eddington luminosity) and galaxy mass of $10^9-10^{11} \msun$.

AGN can be identified via color-color selection \citep{2018NatAs...2..987B,2017ApJ...838..117N,2017ApJ...849..155V,2018MNRAS.476..407V,2022arXiv220714265T,2022arXiv220802822G}, but the success of this technique also depends by how much an AGN inside a star forming galaxy contributes to the total emission. In Fig.~\ref{fig:combcolorstest} we show an example which avoids the bluest filter where high-z sources suffer from intergalactic absorption, but at the cost of adding a MIRI band, less-sensitive than NIRCAM. We highlight that MBHs on the nominal $\log(M_{\rm gal}/\msun)-4$ relation do not contribute enough to the emission to appreciably change the colors from galaxy-dominated to AGN-dominated. MBHs must have mass in excess of  $\log(M_{\rm gal}/\msun)-3$ in order to be generically identifiable, although a fraction of MBHs with mass $\log(M_{\rm gal}/\msun)-3$ and $\log(f_{\rm Edd})=0$ straddle the AGN and galaxy regions, but they are sufficiently separated from the galaxy region to stand out. As a warning, these results assume no attenuation based on the limited extinction in these galaxies: dust in the galaxies would make the galaxy colors redder, moving towards the AGN region, while the AGN emission itself would be reddened by dust in the interstellar medium and in the its vicinity, e.g., in a torus.

\begin{figure}
	\includegraphics[width=\columnwidth]{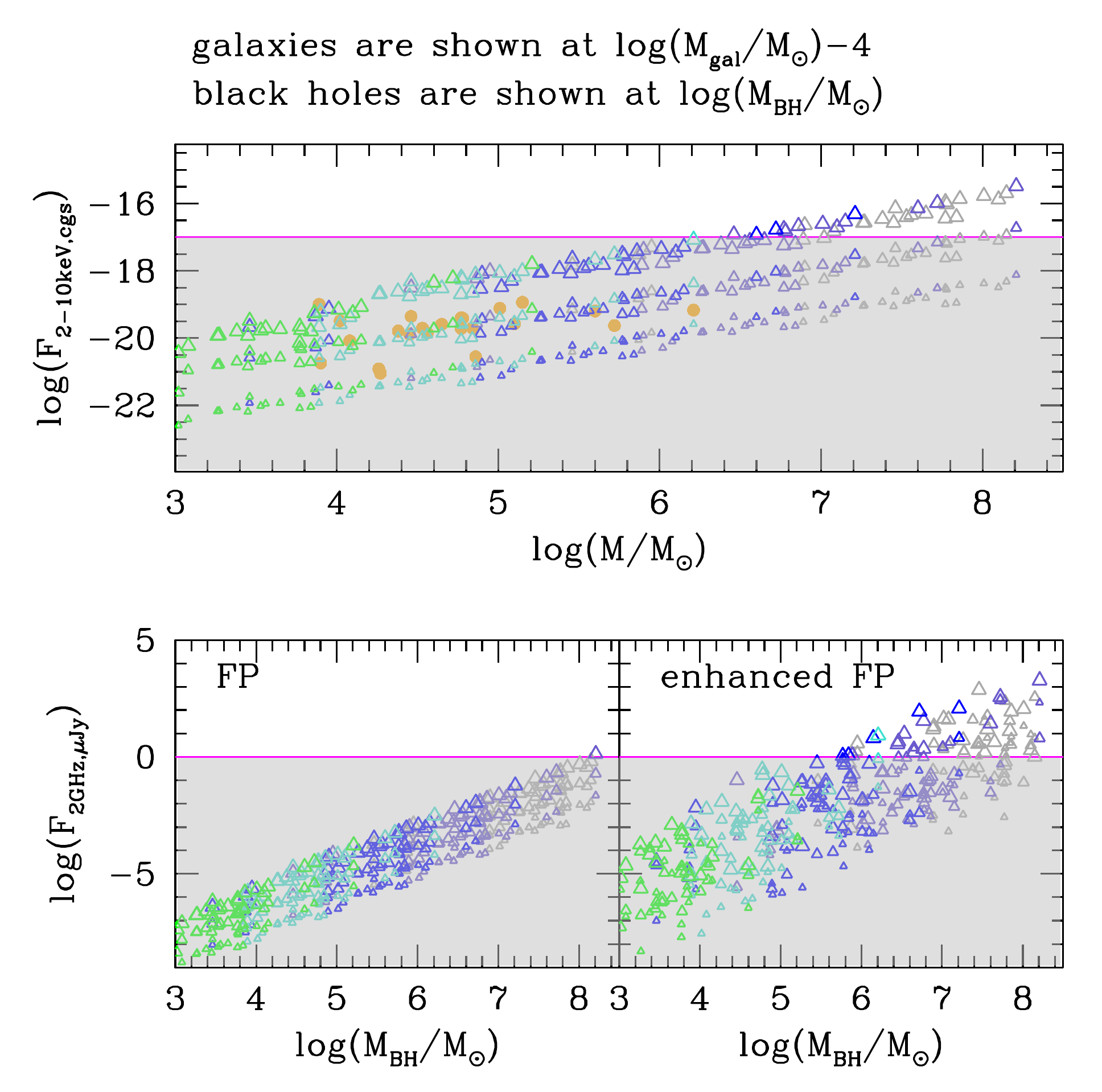}
    \caption{Predicted X-ray and radio fluxes. X-ray luminosity is obtained by integrating the spectrum in the 2-10 keV range (observer's frame) and the radio at 2~GHz luminosity from the fundamental plane.  Green: MBHs with mass $\log(M_{\rm gal}/\msun)-5$; turquoise: MBHs with mass $\log(M_{\rm gal}/\msun)-4$; blue: MBHs with mass $\log(M_{\rm gal}/\msun)-3$; slate grey: MBHs with mass $\log(M_{\rm gal}/\msun)-2$; grey: MBHs with mass $\log(M_{\rm gal}/\msun)-1$.  The size of the symbol scales with the Eddington ratio: small for $\log(f_{\rm Edd})=-2$, medium for $\log(f_{\rm Edd})=-1$, large for $\log(f_{\rm Edd})=0$. The orange dots show the brightness of the X-ray binary population in the host galaxies.   
     The horizontal magenta lines show approximate flux limits for future and upcoming missions (AXIS, Athena, eVLA, SKA). The shaded grey area therefore shows the AGN which remain invisible.}
    \label{fig:Xradio}
\end{figure}

\subsection{Properties or detections of MBHs in X-ray and radio}
The prospect of detecting MBHs hidden in these galaxies is brighter in X-ray and radio, at least under optimistic assumptions (Fig.~\ref{fig:Xradio}).  With the model adopted here at least some MBHs with mass $\log(M_{\rm gal}/\msun)-3$ can be detected. We also find that the X-ray binary flux is below $10^{-18}$ in cgs units in all cases. Empirical models from low-redshift observations predict an X-ray emission from binaries which varies over more than one order of magnitude. This does not affect our conclusions as we have used here a conservative model predicting among the highest emission from binaries, and this remains still well below possible detection by future missions. We remind the reader that the scaling of these models with redshift has yet to be verified at $z>9$. 

In radio the standard FP has no detection except for the most overmassive, and massive, MBHs, while the enhanced FP allows for detections of `normal' MBHs down to the nominal relation used here, MBH mass $\log(M_{\rm gal}/\msun)-4$, although in a very small fraction of cases.

\subsection{Prospects for identifying and understanding MBHs in high-z galaxies}
\label{sec:prospects}

Based on the comparison of MBH and galaxy emission shown in Fig.~\ref{fig:compareL} and Fig.~\ref{fig:compareLzoom}, MBHs are squarely more luminous than galaxies only when they are overmassive and accreting at high Eddington fractions. Some `normal' (not overmassive) MBHs pop up above the galaxy luminosity in the redder JWST bands (e.g., at F2100W), in X-rays and they could be detectable under optimistic assumptions in radio. `Intermediate' mass black holes, with mass less than $10^5 \msun$ are always hidden by the starlight of the host galaxy in JWST bands, as already noted in \cite{2022arXiv220802822G}, and they are simply too faint to be detected in the X-ray and radio bands. 

These examples imply that if MBHs are detected using JWST in this type of galaxies and at these redshifts, they must necessarily be more massive than the relation with galaxy mass implies. We further note that actual MBH mass measurements or estimates via broad lines is hard, if not impossible, for these objects. The bottom line is that if one detects an over-massive black hole they have to be careful in assessing whether this is a selection bias -- the only MBHs that \textit{can} be detected -- or they are representative of the whole population.

\section{What does this mean for seeding models?}\label{seeds}

What if  sufficiently massive MBHs ($>10^6 \msun$ based on the discussion in Section~\ref{multilambda}) are eventually detected in these candidate galaxies above redshift $z=9$? Could we constrain seed models?

An academic exercise to have order of magnitude estimates is to invert the MBH growth rate, starting for example with a $10^6 \msun$ MBH at $z=10$. This MBH had a mass of $10^4 \msun$ at $z\sim 16$ and $10^3 \msun$ at $z\sim 25$, assuming constant accretion at $\log(f_{\rm Edd})=0$. This assumption gives the maximal growth, i.e., the minimal MBH mass that can grow to $10^6 \msun$ MBH at $z=10$. A barely detectable  $10^5 \msun$ MBH at $z=10$ had a mass of $10^4 \msun$ at $z\sim 12$ and $10^3 \msun$ at $z\sim 16$, again assuming constant $\log(f_{\rm Edd})=0$.  Therefore detecting a $10^6 \msun$ MBH in one of the candidate galaxies would suggest the formation  of seeds of $10^3-10^4 \msun$, or super-Eddington accretion, provided it is extended for sufficiently long times \citep{2016MNRAS.456.2993L,2022arXiv220410330S,2022arXiv220108766M,2022ApJ...935..140H}. Lighter seeds, relics of population III stars, are unlikely to have grown at all under Eddington limited accretion \citep{2018MNRAS.480.3762S}, therefore it is even harder to justify their super-Eddington growth.   

A $10^6 \msun$ MBH in one of the candidate galaxies would be over-massive and it may point to `obese' MBHs caused by {\it heavy} seed formation, with its AGN feedback preventing the host galaxy from growing and thus maintaining the `overmassiveness' of the MBHs \citep{2013MNRAS.432.3438A,2018ApJ...865L...9V}. Such detection would suggest that heavy seeds have been formed, but we can not exclude the existence of light seeds, since they would have not grown enough to be detectable.  \cite{2022arXiv220813582O} propose that GL-z12-1, with a mass of $2\times 10^8 \msun$ may be hosting an AGN powered by a MBH with mass $>10^6 \msun$: this is a case of a MBH in a low-mass galaxy where MBH growth is expected to be limited because of the effect of supernova (SN) feedback \citep{2015MNRAS.452.1502D,2017MNRAS.468.3935H,2017MNRAS.472L.109A}. In this case, the MBH should have formed with mass already close to the presumed mass. If we don't detect overmassive/obese MBHs in these galaxies, the conclusion is that heavy seeds do not form in galaxies with  properties similar to those  listed in Table \ref{TableGal}.

Can instead the non-detection of MBHs in these candidate galaxies be used to constrain seed models? In the case of lighter seeds such as  those forming through  dynamical channels \citep[e.g.,][]{2004Natur.428..724P,Frietag2006,2012ApJ...755...81M,2017MNRAS.467.4180S,Boekholt2018,2022MNRAS.512.6192S}, both the age of the stars and the age of the black hole are to be comparable, since the seed can only form in presence of a substantial population of massive stars. The masses of the candidate galaxies and the ages of the stellar populations of 10-100 Myrs, as well as their proposed compactness \citep{2022arXiv220813582O} could be  consistent with formation scenarios of black holes of $\sim 10^3\msun$, formed from runaway collisions of massive stars in young, dense star clusters, with relaxation and collision times as short as a few Myrs \citep{Devecchi2012}. These  MBH would however be  too faint to be detectable. 

\begin{figure*}
\includegraphics[width=\columnwidth]{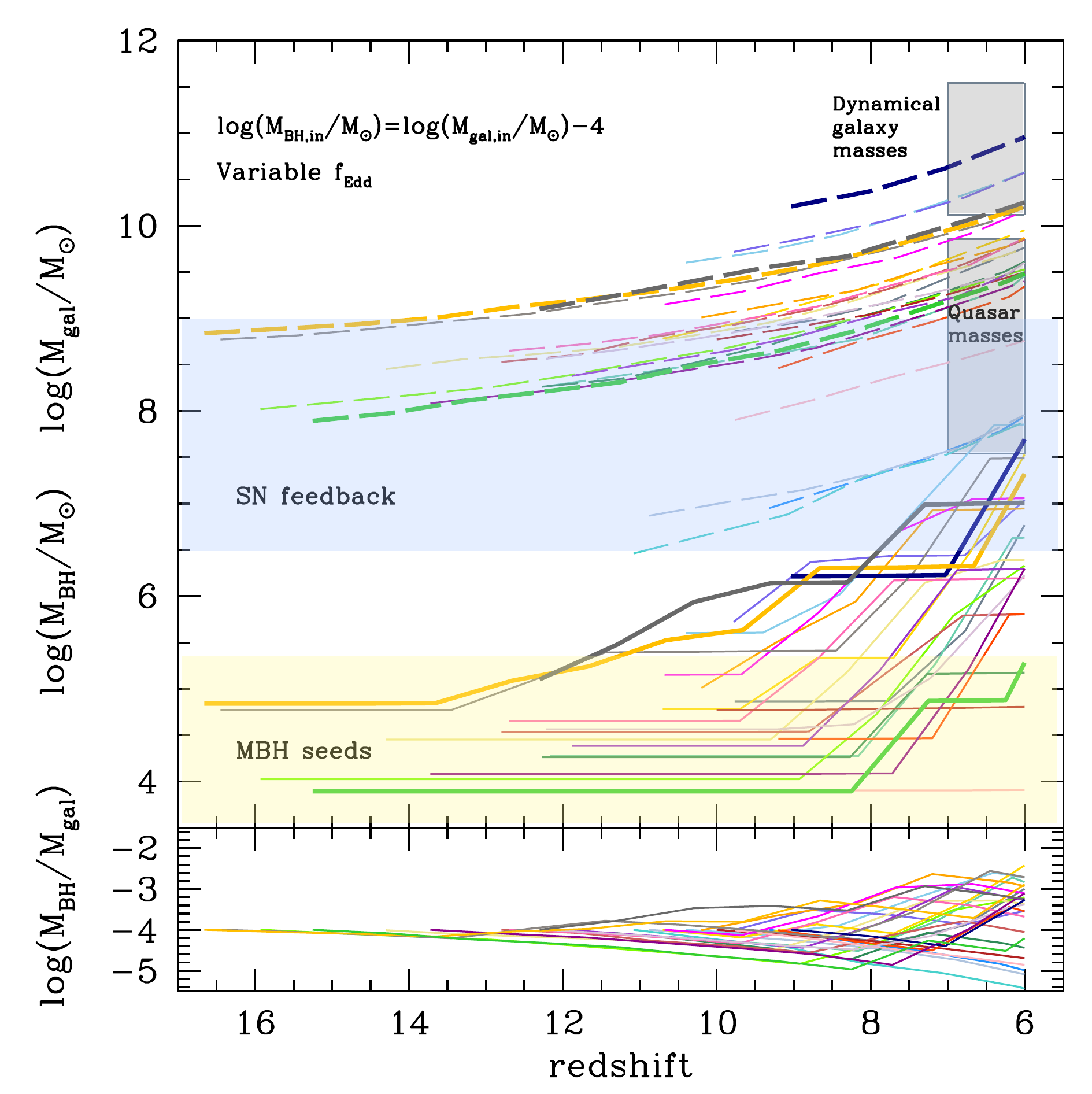}
\includegraphics[width=\columnwidth]{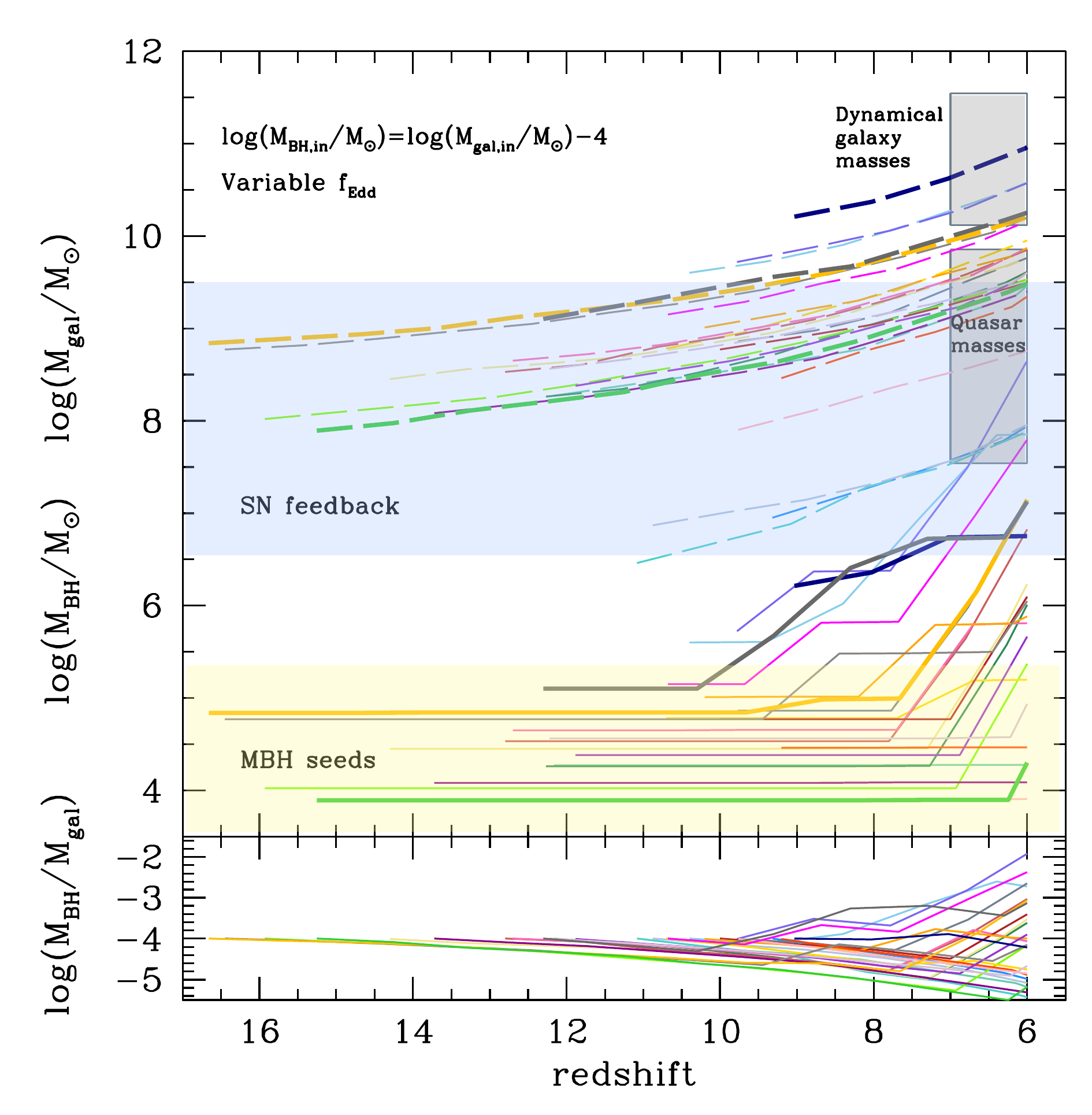}
\includegraphics[width=\columnwidth]{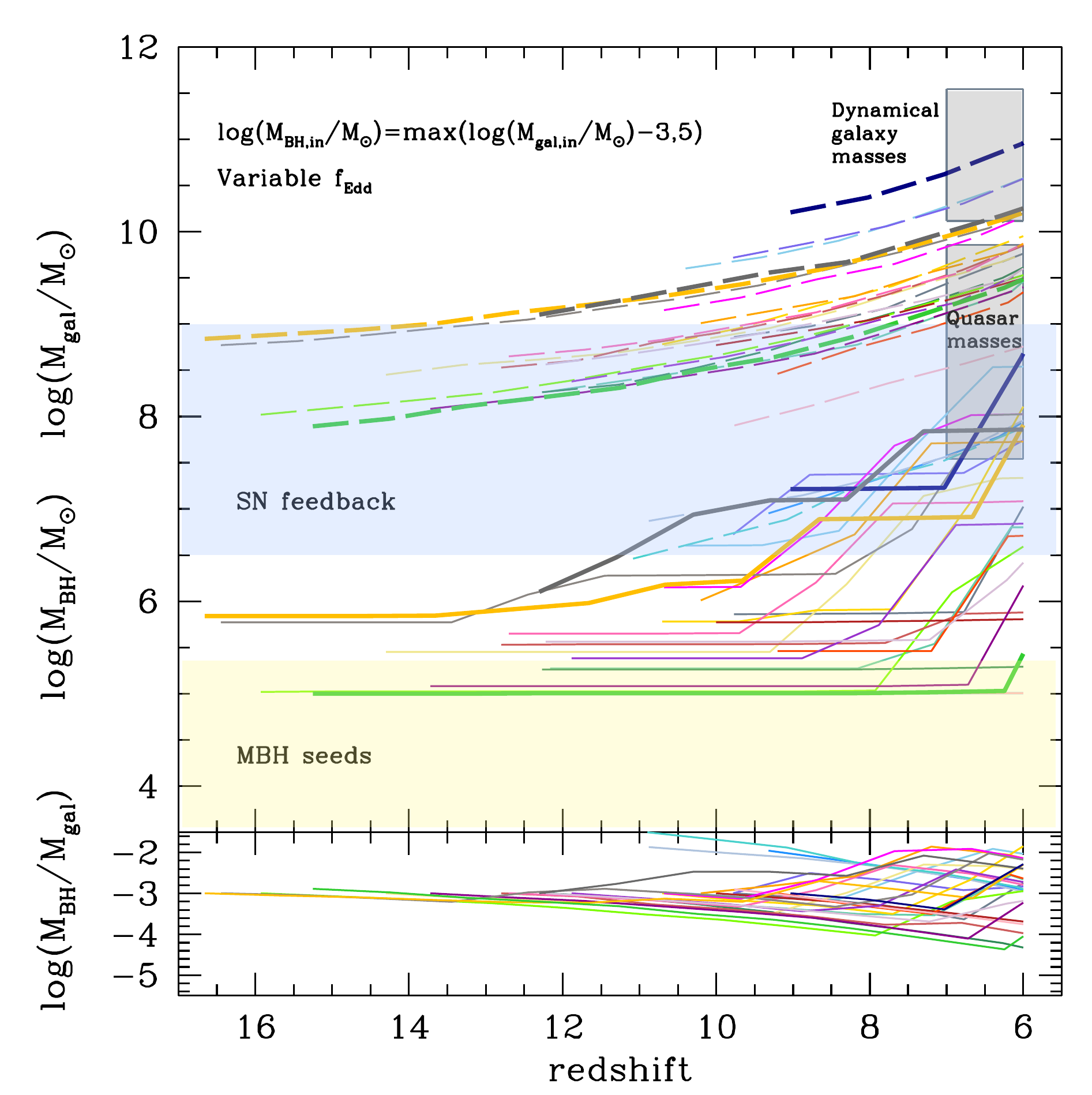}
\includegraphics[width=\columnwidth]{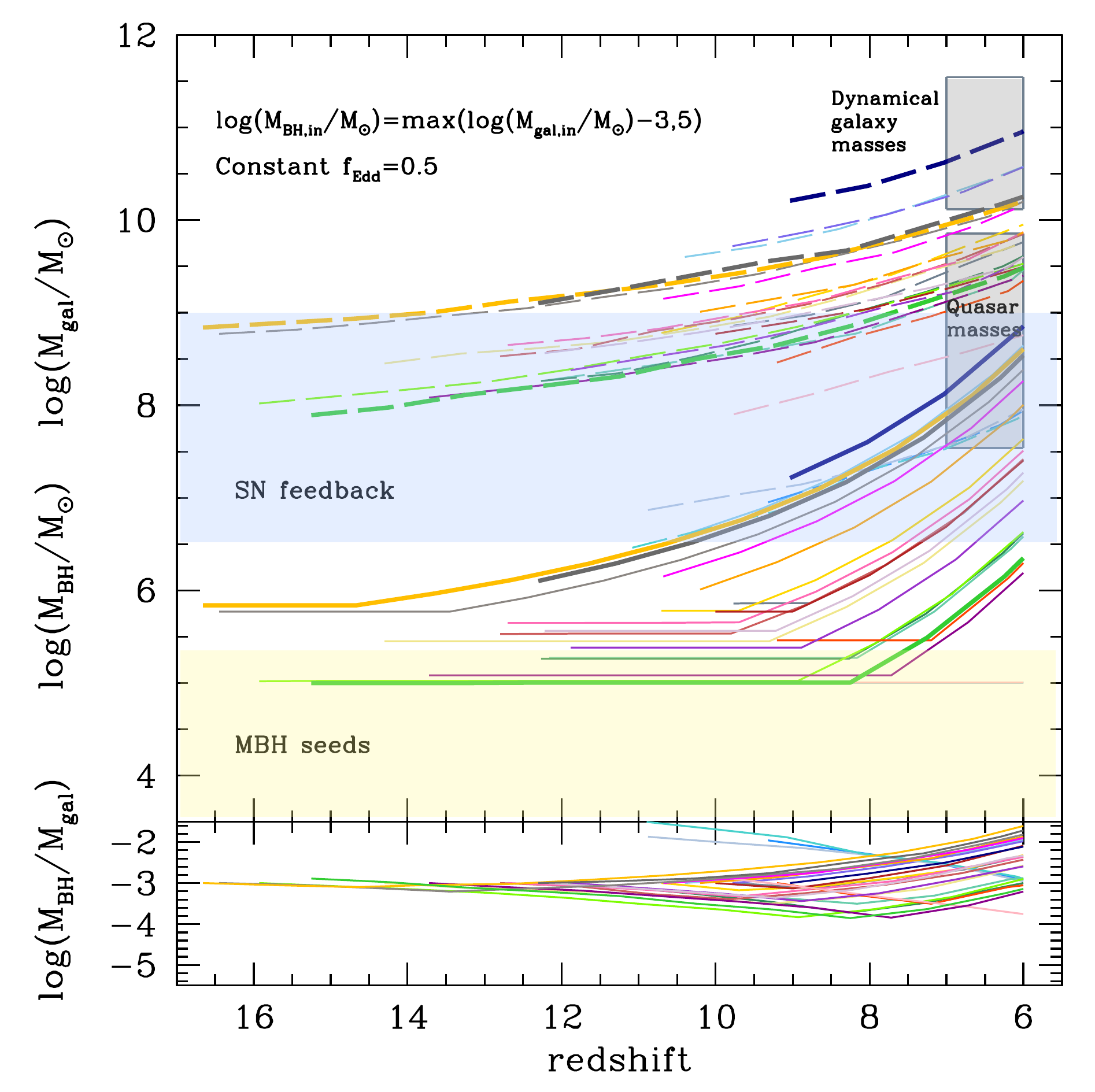}
\caption{Examples of MBH (solid curves) and galaxy (dashed curves) evolution, taking as initial conditions the galaxies in Table 1 and making different assumptions for the MBHs they host (top: $\log(M_{\rm BH}/\msun)=\log(M_{\rm gal}/\msun)-4$; bottom:  $\log(M_{\rm BH}/\msun)=\max(5,\log(M_{\rm gal}/\msun)-3)$), effect of SNae (top-left and bottom: $M_{\rm SN}=10^9 \msun$; top-right: $M_{\rm SN}=3\times10^9 \msun$), and their growth rate (top, bottom-left: scaling with SFR; bottom-right: $f_{\rm Edd}=0.5$). Galaxies are always assumed to grow along the SFR-galaxy mass sequence. The thicker curves highlight some example galaxies: SMACS\_z10b, SMACS\_z16b, S5-z17-1, GL-z12. In all panels the bottom region shows the ratio of MBH to galaxy mass, without highlighting specific galaxies.}
\label{fig:growth}
\end{figure*}

\section{Implications for $z\sim 6-7$ quasars}\label{qsos}

Can we shift the $z=6$ `quasar' problem to a `galaxy' problem at $z>9$? Can these galaxies be progenitors of $z=6$ quasars? 
Let's take as an example GL-z10  and GL-z12 \citep{2022arXiv220709434N}.  The stellar masses are reported to be $\log(M_{\rm gal}/\msun)=9.6$ and $\log(M_{\rm gal}/\msun)=9.1$ respectively and the corresponding number density (given the survey volume) is $4.5\times 10^3$~Gpc$^{-3}$, which is about 3 orders of magnitude larger than the number density of luminous quasars at $z\sim 6-7$. So, of these exceptional young galaxies, only about one in a thousand is needed to be a progenitor of a $z\sim 6-7$ quasar. In practice, the `galaxy' problem at $z>9$ is mostly an abundance issue -- `too many' massive galaxies found in small volumes \citep{2022arXiv220712474F} -- while the $z=6$ `quasar' problem is a mass/timing issue -- not enough time to build up the MBH mass\footnote{ For some high-redshift galaxy candidates, such as those presented in \cite{2022arXiv220712446L} there is also a mass/timing issue. We do not include these candidates in this analysis because, contrary to the other candidates presented in Table~\ref{TableGal} we need to include dust attenuation in order to reproduce their photometry and this adds an additional level of uncertainty in modeling the MBH SEDs as well. \cite{2022arXiv220814999E} suggest that one of these galaxies could be an AGN.}.

The mass of galaxies such as GL-z10 and SMACS\_z10c at the time of observation are already sufficiently massive that SN feedback should not hinder MBH growth. Assuming that these candidate galaxies host an MBH with mass $\log(M_{\rm gal})-4$, this leads to $\log(M_{\rm BH}/\msun)=5.6$ and $\log(M_{\rm BH}/\msun)=5.7$. If the galaxies have a sufficiently regular morphology, MBHs of such mass should be able to remain close to the center \citep{2019MNRAS.486..101P,2021MNRAS.508.1973M}, and there is no reason to expect that such MBHs cannot grow. In the case of $z\sim 16$ galaxies, such as SMACS\_z16b, SMACS\_z16a, CR2-z17-1 and S5-z17-1, their masses are below the threshold where SN feedback stunts MBH accretion, therefore even if they hosted MBHs, the growth of such MBHs would be limited until the galaxies would grow further.

We can consider a simple joint model for the galaxy candidates of Table~\ref{TableGal} and their hypothetical MBHs. Let us assume that the stellar mass of the galaxy increases following the SFR-$M_{\rm gal}$ sequence \citep[e.g.,][]{2022arXiv220711135L}, and that the MBH growth is also modulated by the galaxy growth. We consider here two mass thresholds for $M_{\rm SN}$, $10^9 \msun$ and $3\times 10^9 \msun$, for SN-stunted MBH growth, although in some models this value can be in excess of $10^{10} \msun$ \citep{2022MNRAS.511.5756T}. In galaxies with mass above this threshold we assume the Eddington ratio corresponding to a given fraction of the SFR, and we assume a fraction $10^{-4}$ to be in line with what we consider the `normal' ratio between MBH and galaxy mass. We also consider that this ratio is not constant, but only constant when averaged over long timescales \citep{2014ApJ...782....9H}, and therefore add a Gaussian scatter centered on zero and with $\sigma=1$, but limiting $f_{\rm Edd}$ between 0 and 1. We consider steps of 0.5 in redshift, at each step the galaxy mass gives the SFR, and from the SFR we calculate the Eddington rate, with this information we update the masses and iterate until we reach the final redshift.

We note that if used a constant ratio between MBH accretion rate and SFR the observed properties of the quasars would not be reproduced, since in general the mass ratio for the quasars is higher than `normal', although there could be differences between stellar and dynamical mass \citep{2019MNRAS.488.4004L} and the differences decrease for fainter quasars \citep{2022MNRAS.511.3751H,2019PASJ...71..111I}. Adding scatter in the MBH growth rate allows for some phases of rapid growth, and while for the whole population the average mass ratio between galaxy and mass remains at the assumed level, some MBHs can grow more efficiently. We recall that given the number densities, only about 1 in a 1000 of galaxies like these candidates could have successful joint MBH and galaxy growth to explain the quasar properties. We also consider a simple case with constant $f_{\rm Edd}=0.5$. The initial MBH mass in a somewhat pessimistic case set to $\log(M_{\rm BH}/\msun)=\log(M_{\rm gal}/\msun)-4$ and in an optimistic case to $\log(M_{\rm BH}/\msun)=\max(5,\log(M_{\rm gal}/\msun)-3)$. To estimate which galaxies are most likely to grow and to host a MBH that grows, we run 50 different realizations. 

The results are reported in Fig.~\ref{fig:growth} for one example realization. Most of these candidate galaxies barely reach the range of dynamical masses of the quasar host galaxies \citep{2019PASJ...71..111I,2021ApJ...911..141N}. About 25 per cent of the candidates can reach  $\log(M_{\rm gal}/\msun)>10.3$ by $z=6$, with SMACS\_z10b, GL-z10, S5-z17-1, CR2-z17-1 the most likely cases. 

The top panel shows two conservative cases. Starting with $\log(M_{\rm BH}/\msun)=\log(M_{\rm gal}/\msun)-4$, most MBHs in these candidate galaxies would not grow to the masses of $z\sim 6-7$ quasars.  About 7 per cent of galaxies can host $\log(M_{\rm BH}/\msun)>7.5$ by $z=6$. SMACS\_z10b, GL-z10, S5-z17-1, CR2-z17-1,  and SMACS\_z10c
are the galaxies most likely to host such MBHs. The fraction decreases to 5 per cent if $M_{\rm SN}=3\times 10^9 \msun$ as threshold for MBH growth (Fig.~\ref{fig:growth}, top-left panel). In this case the most likely galaxies are SMACS\_z10b, SMACS\_z10c, GL-z10, GL-z12.

The bottom panel shows two optimistic cases. Starting with $\log(M_{\rm BH}/\msun)=\max(5,\log(M_{\rm gal}/\msun)-3)$, on average 24 per cent of the candidates we study reach quasar-like masses, if $M_{\rm SN}=10^9 \msun$.  Under these more optimistic assumptions,  SMACS\_z10b, GL-z10,  SMACS\_z10c,  GL-z12, S5-z17-1, GL-z9-1, CR2-z17-1, WHL0137-5347 are the most likely to host MBHs with $\log(M_{\rm BH}/\msun)>7.5$ by $z=6$. The fraction decreases to about 15 per cent for $M_{\rm SN}=3\times10^9 \msun$. With $M_{\rm SN}=10^9 \msun$ and fixed $f_{\rm Edd}=0.5$ (Fig.~\ref{fig:growth}, bottom-left panel) almost 43 per cent of the MBHs would reach quasar-like masses, while, for a comparison, with an initial mass $\log(M_{\rm BH}/\msun)=\log(M_{\rm gal}/\msun)-4$ only about 19 per cent of MBHs enter the region with the same assumption on the accretion rate. With $M_{\rm SN}=3\times 10^9 \msun$ the fractions change to 27 and 12 per cent. Finally, if we assumed $M_{\rm SN}=10^{10} \msun$ even the most optimistic scenarios would give no more than 10 per cent of MBHs with mass $\log(M_{\rm BH}/\msun)>7.5$ by $z=6$. 

Here we have not focused specifically on the most massive MBHs powering high-redshift bright quasars: if we required $\log(M_{\rm BH}/\msun)>9$ by $z=6$, we would obtain less than 1 per cent successful cases, under favourable/optimistic assumptions ($\log(M_{\rm BH}/\msun)=\max(5,\log(M_{\rm gal}/\msun)-3)$ and $M_{\rm SN}=10^{9} \msun$). We refer the reader to \cite{2022MNRAS.509.1885P} for a statistical analysis showing the permitted parameter space in seed masses, average Eddington ratios, duty cycles and radiative efficiencies required to produce MBHs with mass $>10^9 \msun$ as a function of redshift.

In summary, some of the candidate galaxies in Table~\ref{TableGal} have reasonable properties for putative `normal' MBHs, which are nevertheless invisible at the time of observation of the candidate galaxies at $z \gtrsim 9$ as shown in Section~\ref{sec:prospects}, to develop into faint $z\sim 6$ quasars, and the different number densities allow  for about only 1 in 1000 to need to develop this way. These MBHs must have a  mass of $\sim 10^4-10^5 \msun$ at $z\sim 10-16$. 

In candidate galaxies with masses $<M_{\rm SN}$, MBH growth is inefficient, therefore for this to work the MBH mass at birth \emph{must} have been already close to $\sim 10^6\msun$.  Lighter seeds, say $\sim 10^3 \msun$, \emph{must} have formed in galaxies that reached  $M_{\rm SN}$ at an earlier redshift and with enough gas supply to have remained on the SFR-$M_{\rm gal}$ the whole time. We also speculate that the compactness of many of high-redshift galaxies  \cite{2022arXiv220813582O} could favor MBH growth and help them grow close to the Eddington limit \citep{2019MNRAS.484.4413H}. 

\cite{2022arXiv220714808M} suggest that the candidate galaxies discussed in this paper are exceptional only in being very young, besides having little dust and being more numerous than expected \citep[e.g.,][]{2022arXiv220709434N,2022arXiv220712474F}. Therefore in principle the progenitors of the high-$z$ quasars may have developed even earlier, producing more massive and perhaps dustier galaxies that have not been detected yet at such redshifts \citep[but see][]{2022arXiv220712446L}, making the growth of the MBH from a small seed in such galaxies less challenging, and we have not seen these galaxies yet.

\section{Conclusions}

We have investigated what type of MBHs would be detectable in $z \geqslant 9$ galaxies, if they are as young and star-forming as the galaxy candidates presented in Table~\ref{TableGal}. We have also explored what the detection or non-detection of MBHs/AGN in this type of galaxies implies for MBH seed models and for the progenitors of $z>6$ quasars. We summarize in the following our results.

\begin{itemize}

\item MBHs with a mass that scales with galaxy mass as at $z=0$ or expected at high-redshift based on empirical data-driven models \citep{2021arXiv210510474Z}  would be significantly fainter than the stellar component. Only `overmassive' MBHs have a chance of being detected via color-color selection or by outshining the host galaxy in JWST bands. The situation is similar with X-ray and radio observations, unless the radio emission is enhanced with respect to standard expectations. 

\item Some among the high-redshift candidates could be reasonable cradles for `normal' MBHs, which are hidden from view at the time of the observation, to develop into $z\sim 6$ quasars. The rarity of $z\sim 6$ quasars with respect to these candidates is such that only about 1 in 1000 needs to grow their MBHs fast. 

\item Only `overmassive' MBHs can be detected in this type of galaxies: this means that detections have to be treated with care: are MBHs really `overmassive' or are most MBHs `normal' and we simply cannot detect them? This has important consequence on the interpretation of observation in light of seed models. For some type of models the prediction of `obese' MBHs has to be carefully assessed against the observational bias we have identified. 

\item The masses, SFRs and compactness of galaxies of this type could be conducive to the birth of dynamically-formed seeds close to the time of observation.

\end{itemize}

After submission of this paper, four spectroscopically-confirmed galaxies at $z>10$ have been reported \citep{2022arXiv221204568C,2022arXiv221204480R}. Their properties are fully consistent with the candidates analysed here and the results of our model are unchanged. We have included a figure with results of our model for these galaxies in the Appendix.

JWST surveys can detect many high-redshift AGN \citep{2022arXiv221101389T}, and two candidates have been proposed at $z\sim5$ \citep{2022arXiv220907325O} and at $z\sim 12$ \citep{2022arXiv220813582O}. A statistical sample of AGN will allow to address questions on MBH seeding and galaxy (co)evolution, provided that the hosts are not too bright and star-forming thus outshining the AGN. In this paper we have focused on galaxy candidates that are young and star-forming, but a more varied galaxy population will presumably make it easier to identify AGN, although attenuation will affect the AGN emission. 

Besides detections of MBHs with JWST, constraints on the MBH population at high redshift and seeding models will come from the LISA gravitational wave antenna \citep{LISA2017}.  We briefly speculate here on the properties of these galaxies in relation to the mergers of speculative MBH binaries they could also host. Sersic indices, can be estimated from fitting of JWST high-z candidates photometry and they can inform us on the merging timescale of MBH binaries in these galaxies. \cite{2019MNRAS.487.4985B} estimate that one needs a Sersic index of 3 or greater for a binary of $10^5\msun$ to coalesce in less than 100~Myr, and a Sersic index of 2 for a coalescence in 700~Myr.  \cite{2022arXiv220709434N} and \cite{2022arXiv220712338A} give Sersic index $<1$ for most of their galaxies, although SMACS\_z12b and SMACS\_z16b have 4 and 2.8 respectively. \cite{2022arXiv220813582O} fix the Sersic index at 1.5 based on the trends found in \cite{2015ApJS..219...15S}. With a Sersic of 1.5 a $10^5\msun$ binary would merge in 1~Gyr, which still implies a high-redshift MBH merger detectable by LISA.

\section*{Acknowledgements}
We thank the referee for a constructive and prompt review. MV thanks Raffaella Schneider, Laura Pentericci and Marco Castellano for enlighting discussions. MV also thanks Sapienza University in Rome for kind hospitality. MV thanks Akim Atek for clarifying the results on some candidate galaxies and Chi An Dong-Paez for thoughtful comments on the paper. 

\section*{Data Availability}
Observational data is available in the references provided in Table~\ref{TableGal}. The data underlying this article will be shared on reasonable request to the corresponding author. Most of the results are based on analytical calculations that can be reproduced using the described methodology. 

\bibliographystyle{mn2e}
\bibliography{biblio}
\appendix

\section{Photometry for galaxies with multiple filter data}
\label{App:phot}

\begin{figure}
	\includegraphics[width=\columnwidth]{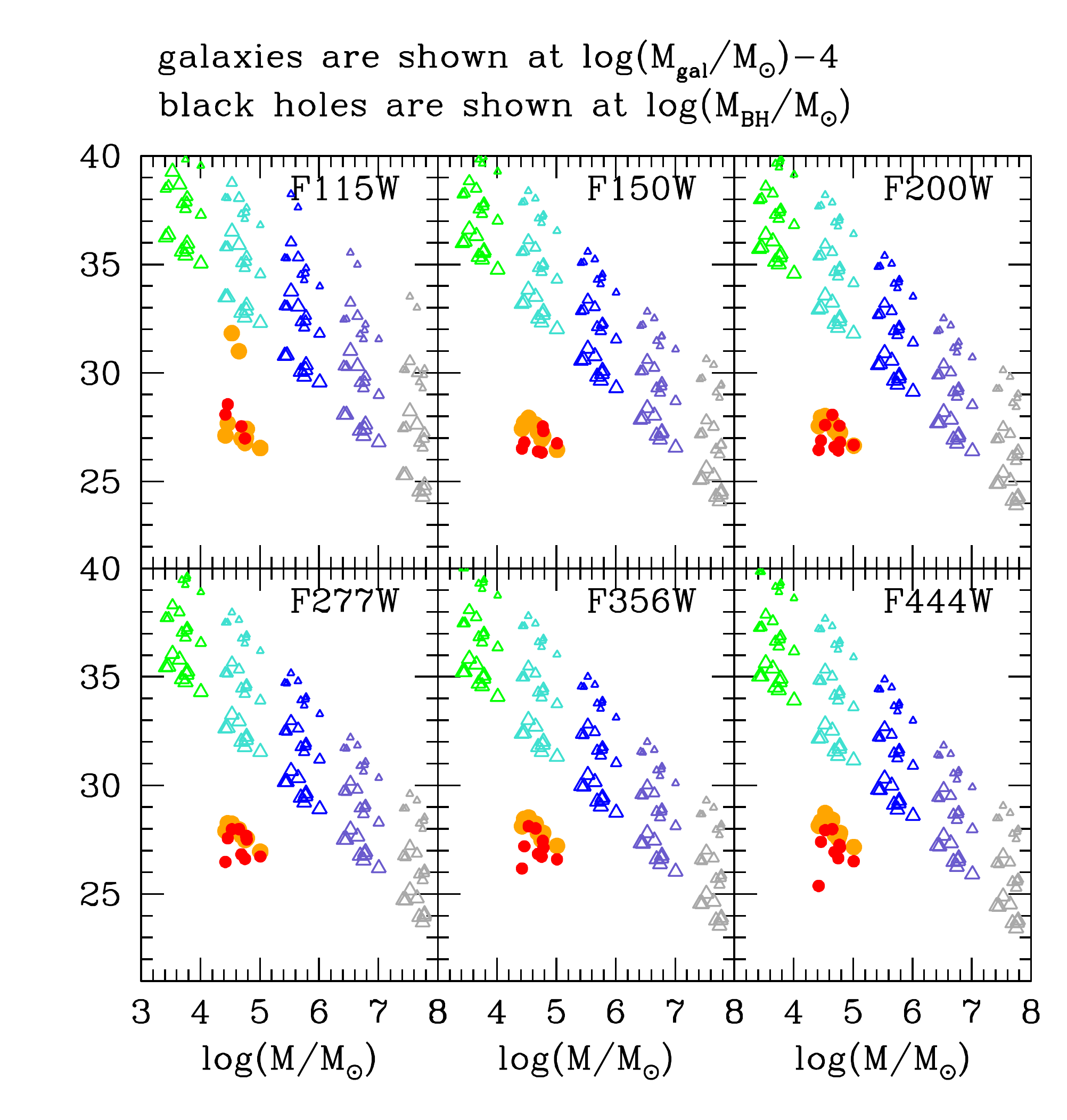}
    \caption{Comparison of AGN and galaxy luminosity using our models for all unlensed galaxies in \citet{2022arXiv221001777B}. 
    Triangles show the AGN. Green: MBHs with mass $\log(M_{\rm gal}/\msun)-5$; turquoise: MBHs with mass $\log(M_{\rm gal}/\msun)-4$; blue: MBHs with mass $\log(M_{\rm gal}/\msun)-3$; slate grey: MBHs with mass $\log(M_{\rm gal}/\msun)-2$; grey: MBHs with mass $\log(M_{\rm gal}/\msun)-1$. The size of the symbol scales with the Eddington ratio: small for $\log(f_{\rm Edd})=-2$, medium for $\log(f_{\rm Edd})=-1$, large for $\log(f_{\rm Edd})=0$. Red dots: photometry from \citet{2022arXiv221001777B}. Orange dots: galaxy apparent magnitude from our models. Yellow circles: galaxy apparent magnitude from our models including dust attenuation.  Red, orange dots and yellow circles are shown at the mass corresponding to $\log(M_{\rm gal}/\msun)-4$. }
    \label{fig:multif}
\end{figure}

\begin{figure}
	\includegraphics[width=\columnwidth]{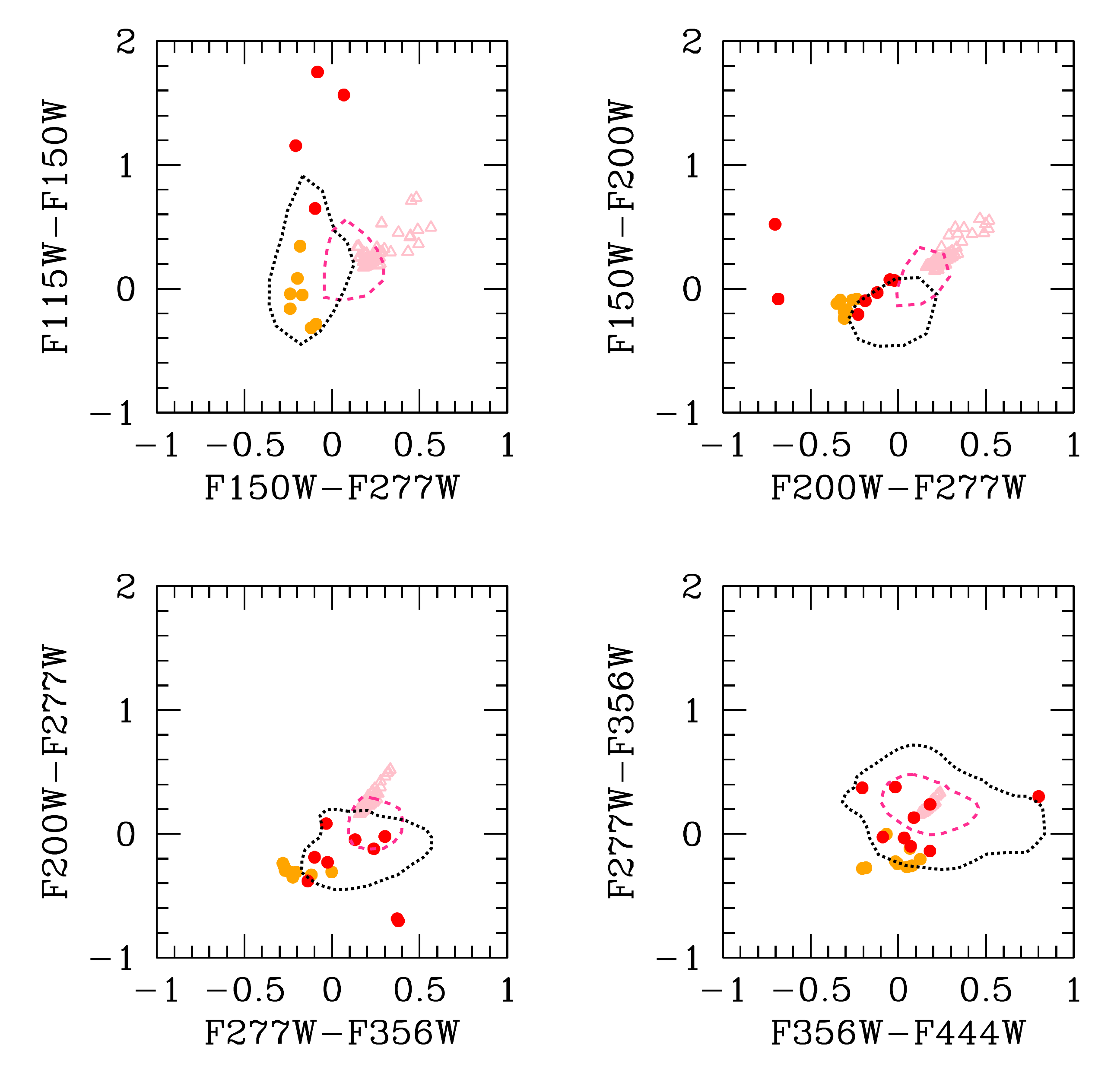}
    \caption{AGN and galaxy colors using our models for all unlensed galaxies in \citet{2022arXiv221001777B}. We compare our models (orange: galaxies; pink: AGN) with (i) the colors obtained from the published photometry (red), and (ii) the regions occupied by galaxies (black dotted contour) and galaxy+AGN composite (magenta dashed contour) in \citet{2022arXiv220802822G}. The magnitudes used to calculate colors in our model include integration only redwards of 912 \AA, therefore short wavelength filters are affected by intergalactic absorption. }
    \label{fig:multif_colcomp}
\end{figure}

In this section we apply our model to all unlensed galaxies in \cite{2022arXiv221001777B}, where 6 photometric points are provided to the reader. Note that the calculation of the magnitude in F115W, and to a lesser extent in F150W, is affected by the Lyman limit: we only integrate the SED (convolved with the filter response) redwards of 912 \AA. In general our model, despite its simplicity, is in good agreement with the photometry (Fig.~\ref{fig:multif}). For this sample, we can confirm that the AGN are fainter than the galaxies in all 6 filters, unless overmassive.  Fig.~\ref{fig:multif_colcomp} compares the colors to the results in \citet{2022arXiv220802822G}, showing reasonable agreement. 

\begin{figure}
	\includegraphics[width=\columnwidth]{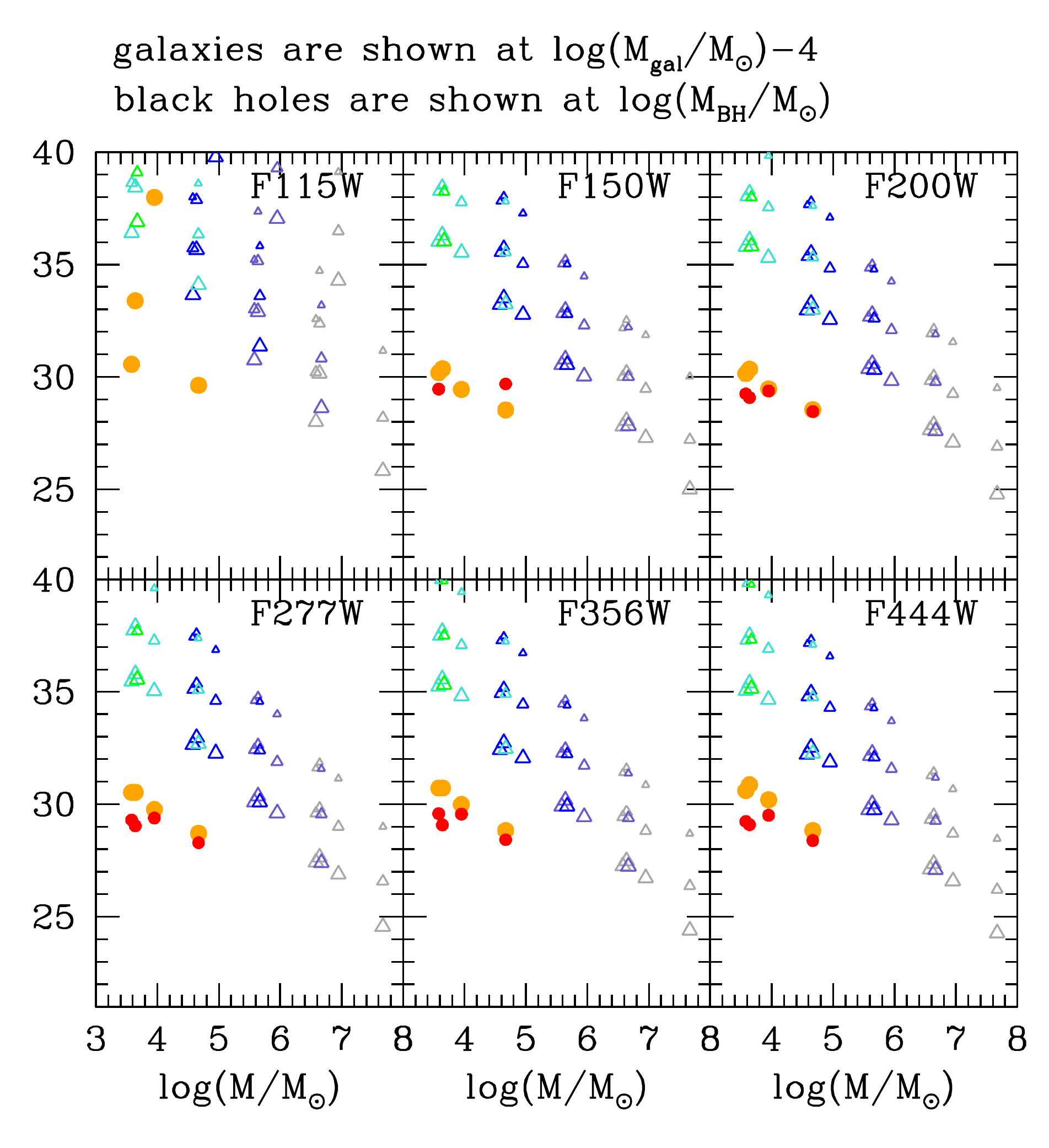}
    \caption{Analogue of Fig.~\ref{fig:multif}, for the four spectroscopically-confirmed galaxies at $z>10$ presented in \citet{2022arXiv221204568C} and \citet{2022arXiv221204480R}.}
    \label{fig:JADES}
\end{figure}

\label{lastpage}

\end{document}